\documentclass[twocolumn,preprintnumbers,amsmath,amssymb,prb]{revtex4}

\usepackage{graphicx}
\usepackage{dcolumn}
\usepackage{bm}
\usepackage{amssymb}
\usepackage{epstopdf}
\usepackage{color}
\usepackage{dsfont}

\begin{document}

\title{Optically induced Lifshitz transition in bilayer graphene}
\author{I. V. Iorsh$^{1}$}
\author{K. Dini$^2$}
\author{O. V. Kibis$^{3}$}\email{Oleg.Kibis(c)nstu.ru}
\author{I. A. Shelykh$^{1,2}$}
\affiliation{$^1$ITMO University, Saint Petersburg 197101, Russia}
\affiliation{$^2$Science Institute, University of Iceland, Dunhagi
3, IS-107, Reykjavik, Iceland}\affiliation{${^3}$Department of
Applied and Theoretical Physics, Novosibirsk State Technical
University, Karl Marx Avenue 20, Novosibirsk 630073, Russia}

 \begin{abstract}
It is shown theoretically that the renormalization of the electron
energy spectrum of bilayer graphene with a strong high-frequency
electromagnetic field (dressing field) results in the Lifshitz
transition --- the abrupt change in the topology of the Fermi
surface near the band edge. This effect substantially depends on
the polarization of the field: The linearly polarized dressing
field induces the Lifshitz transition from the quadruply-connected
Fermi surface to the doubly-connected one, whereas the circularly
polarized field induces the multicritical point, where the four
different Fermi topologies may coexist. As a consequence, the
discussed phenomenon creates physical basis to control the
electronic properties of bilayer graphene with light.
\end{abstract}

\maketitle

\section{Introduction}

In the last decades, the achievements in the laser technology and
microwave techniques have made possible the control of various
structures with a high-frequency electromagnetic field (dressing
field), which is based on the Floquet theory of periodically
driven quantum systems (Floquet
engineering)~\cite{Hanggi_98,Kohler_05,Bukov_15,Holthaus_16}. As a
consequence, the study of condensed-matter structures strongly
coupled to light has become an excited field of modern physics
with the objective to find unique exploitable features.
Particularly, electronic properties of various nanostructures
coupled to a dressing field
--- including quantum wells
\cite{Wagner_10,Teich_13,Kibis_14,Pervishko_15,Dini_16,Morina_15},
quantum rings
\cite{Sigurdsson_14,Kibis_13,Koshelev_15,Joibari_14}, graphene
\cite{Kristinsson_16,Kibis_16,Glazov_14,Perez_14,Syzranov_13,Kibis_11_1,
Kibis_10,Oka_09,Kibis_17}, topological insulators
\cite{Calvo_15,Yudin_16,Torres_14,Usaj_14,Ezawa_2013}, etc --- are
currently in the focus of attention. Developing this scientific
trend in the present paper, we elaborated the theory to control
the topology of the Fermi surface of bilayer graphene (BLG) with a
dressing field.

BLG is the two-dimensional material consisting of two graphene
monolayers, which excite enormous interest of the condensed-matter
community during last years~\cite{bilayer1,AA1}. In contrast to
usual graphene monolayer with the linear (Dirac) electron
dispersion~\cite{CastroNeto_09,DasSarma_11}, the characteristic
electronic properties of BLG are the massive chiral electrons and
the so-called trigonal warping
--- the triangular perturbation of the circular iso-energetic
lines near the band edge. As a consequence, the Fermi surface of
BLG near the band edge consists of several electron ``pockets''
which are very sensitive to external impacts. Particularly, the
gate voltage or uniform mechanical stress crucially changes the
structure of the pockets. This results in the Lifshitz phase
transition
--- the abrupt change in the topology of the Fermi surface~\cite{Falko1,Falko2,Shtyk2017,Dutreix2016}. Although
a strong electromagnetic field is actively studied in last years
as a tool to control electronic properties of BLG --- to induce
the valley currents~\cite{IRbil1,IRbil2}, to create the Floquet
topological insulator~\cite{IRbil3}, to produce additional Dirac
points~\cite{IRbil4}, etc --- the optical control of the Lifshitz
transition in BLG still await for consideration. The present paper
is aimed to fill partially this gap in the theory.

The paper is organized as follows. In Sec. II, we apply the
conventional Floquet theory to derive the effective Hamiltonian
describing stationary properties of electrons in irradiated BLG.
In Sec. III, we discuss the dependence of renormalized electronic
characteristics of the irradiated BLG on parameters of the
dressing field. The last Sec. IV contain the Conclusion and
Acknowledgments.

\section{Model}
\begin{figure}
\includegraphics[width=0.9\columnwidth]{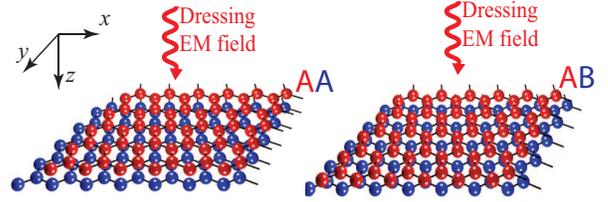}
\caption{\label{fig:figure1} (Color online) Sketch of the system
under consideration: AA-stacked and AB-stacked bilayer graphene
subjected to an electromagnetic wave (dressing EM field)
propagating perpendicularly to the graphene plane.}
\end{figure}

There are the two different crystal structures of BLG, which are
known as a AA-stacked and AB-stacked BLG~\cite{bilayer1,AA1} (see
Fig.~1). Since electronic properties of BLG strongly depends on
the stacking geometry, let us consider these two structures
successively. For the case of the AB-stacked BLG, its low-energy
electronic states are described by the four-band
Hamiltonian~\cite{bilayer1}
\begin{align}\label{01}
\hat{{\cal H}}^\prime_{AB}=\begin{pmatrix} 0 & v_1 \pi^{\dagger} & 0 & v_3 \pi \\
v_1 \pi & 0 & t_1 &  0\\ 0 & t_1 & 0 & v_1 \pi^{\dagger} \\ v_3
\pi^{\dagger} & 0 &v_1 \pi & 0
\end{pmatrix},
\end{align}
where $\pi = \xi p_x + i p_y$, $p_{x,y}$ is the electron momentum
in the BLG plane, $\xi= \pm1$ is the valley index corresponding to
the electron states in the two different $K$-points of the
Brillouin zone, $K_{\xi}=\left({4\xi\pi}/{3 a},0\right)$, $a$ is
the interatomic distance, $ v_{1,3} =\sqrt{3} a t_{1,3}/2 \hbar $
are the characteristic electron velocities, $t_{1}$ and $t_{3}$
are the characteristic energies of the interatomic electron
hopping in the BLG plane and between the two graphene layers,
correspondingly. Eliminating electron orbitals related to dimer
sites, the Hamiltonian (\ref{01}) can be reduced to the two-band
effective Hamiltonian~\cite{bilayer1}
\begin{align}\label{02}
\hat{{\cal H}}_{AB}=-\alpha\begin{pmatrix} 0 &  \left(
\pi^{\dagger}\right)^2 \\  \pi^2  & 0
\end{pmatrix}
+ v_3 \begin{pmatrix} 0 &  \pi\\ \pi^{\dagger} & 0
\end{pmatrix},
\end{align}
where $\alpha=1/2m+v_3 a / 4 \sqrt{3} \hbar$ and $ m = t_1 / 2
v_1^2$ is the effective electron mass. The effective Hamiltonian
(\ref{02}) describes the most interesting low-energy electronic
properties of AB-stacked BLG --- particularly, massive chiral
electrons and trigonal warping \cite{bilayer1,BP} --- and,
therefore, will be used by us in the following analysis. To
describe the interaction between electrons in BLG and a dressing
EM field within the conventional minimal coupling approach, we
have to make the replacement, $\mathbf{p}\rightarrow
\mathbf{p}-e\mathbf{A}$, in the Hamiltonian (\ref{02}), where
$\mathbf{A}=(A_x,A_y)$ is the vector potential of the dressing
field. Assuming the EM wave to be propagating perpendicularly to
the BLG plane (see Fig.~1), the vector potential can be written as
\begin{align}
\mathbf{A}=\frac{E_0}{\omega}(\cos \theta\cos\omega t,\sin
\theta\sin \omega t),
\end{align}
where $\omega$ is the frequency of the EM wave, $E_0$ is the
amplitude of the wave, and the angle $\theta$ defines the
polarization of the EM wave: The angles $\theta=0$ and
$\theta=0\pi/2$ correspond the two linear polarizations, whereas
the angles $\theta=\pi/4$ and $\theta=-\pi/4$ correspond to the
two circular polarizations. Then the time-dependent Hamiltonian
(\ref{02}) can be rewritten as
\begin{align}\label{03}
\hat{{\cal H}}_{AB}(t)=\hat{{\cal
H}}_{0}+\left[\displaystyle\sum_{n=1}^{\infty}\hat{V}_ne^{in\omega
t}+\mathrm{H.c.}\right],
\end{align} where the time-independent part is
\begin{align}
&\hat{{\cal H}}_0=\hat{{\cal H}}_{AB}-\frac{\alpha e^2
E_0^2}{2\omega^2}\begin{pmatrix} 0 & \cos 2\theta \\  \cos 2\theta
& 0
\end{pmatrix},
\end{align}
and the two first harmonics are
\begin{align}
&\hat{V}_1=\frac{\alpha |e| E_0 \sqrt{2}}{\omega}\left[\begin{pmatrix} 0 & p_x\cos(\theta+\xi\pi/4) \\ -p_x\cos(\theta-\xi\pi/4)  & 0 \end{pmatrix}\right.\nonumber\\
&+\left.\begin{pmatrix} 0 & ip_y\sin(\theta-\xi\pi/4) \\ -ip_y\sin(\theta+\xi\pi/4)  & 0 \end{pmatrix}\right]\nonumber\\
&+\frac{v_3|e|E_0}{\sqrt{2}\omega}\begin{pmatrix}0 & \sin(\theta+\xi\pi/4) \\ -\sin(\theta-\xi\pi/4) & 0 \end{pmatrix},\\
&\hat{V}_2=-\frac{\alpha e^2 E_0^2}{2\omega^2}\begin{pmatrix} 0 &
\sin^2(\theta-\xi\pi/4) \\  \sin^2(\theta+\xi\pi/4) & 0
\end{pmatrix}.
\end{align}
Applying the standard Floquet-Magnus approach~\cite{FM1,FM2,FM3}
to renormalize the time-dependent Hamiltonian (\ref{03}) and
restricting the consideration by the terms $\sim1/\omega^2$, we
arrive at the effective time-independent Hamiltonian
\begin{eqnarray}\label{12}
\hat{{\cal H}}_{\mathrm{eff}}&=&\hat{{\cal
H}}_0+\sum_{n=1}^{\infty}\frac{\left[\hat{V}_n,\hat{V}^\dagger_{n}\right]}{\hbar
n\omega}\nonumber\\
&+&\sum_{n=1}^{\infty}\frac{ \left[\left[\hat{V}_n,\hat{{\cal
H}}_0\right],\hat{V}^\dagger_{n}\right]+\mathrm{H.c.}}{2(\hbar
n\omega)^2}.
\end{eqnarray}
To rewrite the Hamiltonian (\ref{12}) in the dimensionless form,
let us introduce the dimensionless electron momentum,
$\mathbf{p}\rightarrow\mathbf{p}/(v_3/\alpha)$, and the
dimensionless electron energy,
$\varepsilon\rightarrow\varepsilon/(v_3^2/\alpha)$. Then the
dimensionless Hamiltonian (\ref{12}) depends only on the two
dimensionless parameters, $\delta_1={|e|E_0\alpha}/{v_3\omega}$
and $\delta_2={v_3^2}/{\alpha\hbar\omega}$. Physically, the first
of them is the ratio of the period-averaged momentum adsorbed by
an electron from the dressing field, $|e|E_0/\omega$, and the
characteristic trigonal-warping momentum, $v_3/\alpha$. The second
one is the ratio of the characteristic trigonal-warping energy,
$v_3^2/\alpha$, and the photon energy, $\hbar\omega$. If the
dressing field is high-frequency enough, the both parameters,
$\delta_1$ and $\delta_2$, are small. Correspondingly, we can
write the Hamiltonian as an expansion on the the small parameter,
$\delta_1\delta_2\ll1$. Then the dimensionless Hamiltonian
(\ref{12}) in the vicinity of the $K_\xi$-point ($p<1$) reads
\begin{align}\label{Heff}
\hat{{\cal H}}_{\mathrm{eff}}=\hat{{\cal
H}}_{AB}+\mathbf{g}(\mathbf{p}){\bm{\sigma}},
\end{align}
where $\bm{\sigma}=(\sigma_x,\sigma_y,\sigma_z)$ is the Pauli
matrix vector, and $\mathbf{g}(\mathbf{p})=(g_x,g_y,g_z)$ is the
vector with the components
\begin{align}
&g_x={(\delta_1\delta_2)^2}[p_y^2[\cos^2\theta-3-4\xi
p_x(4\cos^2\theta-1)-4p^2\cos^2\theta]\nonumber\\
&+p_x^2(5-8\xi p_x+4p^2)\sin^2\theta-\xi
p_x\sin^2\theta]-\frac{\delta_1^2}{2}\cos2\theta,\nonumber\\
&g_y=\frac{(\delta_1\delta_2)^2}{2}[2p_y[\cos^2\theta+2p^2(3-2\cos^2\theta)]\nonumber\\
&+2\xi p_xp_y(2\cos^2\theta+6+12\xi p_x\cos 2\theta+4p^2)],\nonumber\\
&g_z=\frac{\xi\delta_1^2\delta_2}{2}(1-4p^2){\sin2\theta}\nonumber.
\end{align}

As to the case of the AA-stacked BLG, its low-energy electronic
states are described by the four-band Hamiltonian~\cite{AA1}

\begin{align}\label{AAHD}
\hat{{\cal H}}^\prime_{AA}=&\begin{pmatrix} 0 & v_0 \pi^{\dagger}
& t_1 & 0 \\ v_0 \pi & 0 &0 &  t_1 \\ t_1 & 0 & 0 & v_0
\pi^{\dagger}
\\0 &  t_1 &v_0 \pi & 0
\end{pmatrix},
\end{align}
where $t_1$ is the interlayer hopping constant and $v_0$ is the
electron velocity in monolayer graphene. Diagonalizing the
Hamiltonian (\ref{AAHD}), we arrive at the electron energy
spectrum,
\begin{align}
\varepsilon=\pm t_1\pm v_0p,
\end{align}
which exhibits the two monolayer graphene dispersions,
$\varepsilon=\pm v_0p$, shifted with respect to each other by the
twice interlayer hopping constant, $2t_1$. It should be stressed
that the interlayer hopping direction of the AA-stacked BLG is
orthogonal to the polarization vector of the normally incident
dressing field. As a consequence, the dressing field does not
affect the interlayer coupling in BLG and renormalizes electronic
properties of each monolayer independently. Therefore, the effect
of the dressing field on the AA-stacked BLG can be reduced to the
known problem of dressed graphene monolayer~\cite{Kristinsson_16}.
Since the electromagnetic dressing of graphene monolayer does not
lead to the Lifshitz transition, we will focus the following
analysis only on the AB-stacked BLG.

\section{Results and Discussion}

If the dressing field is linearly polarized along the $x$ axis
($\theta=0$), the Hamiltonian (\ref{Heff}) reads
\begin{eqnarray}\label{HL}
\hat{{\cal H}}_{\mathrm{eff}}&=&{\sigma}_x[\xi p_x+p_y^2[1-(\delta_1\delta_2)^2(2+12p_x+4p^2)]-{\delta_1^2}/{2}\nonumber\\
&-&p_x^2]-{\sigma}_y[p_y(1-(\delta_1\delta_2)^2[1+2p^2])+2\xi p_x p_y\nonumber\\
&\times&(1-(\delta_1\delta_2)^2[4+6\xi
p_x+4p^2])]+O(\delta_1^4\delta_2^4).\label{heffLin}
\end{eqnarray}
Diagonalizing the Hamiltonian (\ref{HL}), we arrive at the energy
spectrum of dressed electrons near the $K_\pm$ point, which is
plotted in Fig.~2. In the absence of irradiation, the equi-energy
map shows the four electron pockets near the $K_\pm$ point, which
consist of one central pocket and three ``leg'' pockets (see
Fig.~2a). These four pockets are positioned at the momenta
$\mathbf{p}=0$ and $\mathbf{p}=(\xi\cos(2n\pi/3),\sin(2n\pi/3))$,
where $n=0,1,2$. The dressing field distorts the pockets, shifting
their position in the Brillouin zone (see Fig.~2b). Taking into
account only the leading term of the Hamiltonian~\eqref{heffLin},
we arrive at the equation,
$$
(\delta_1^2/2+p_x^2-p_y^2-\xi p_x)^2+(p_y+2\xi p_xp_y)^2=0,
$$
describing the centers of the shifted pockets, $p_x$ and $p_y$.
This equation has the four solutions: The two ones correspond to
the momenta $p_y=0$ and $p_x=(\xi\pm\sqrt{1-2\delta_1^2})/2$,
whereas the other two solutions are
$\mathbf{p}=(-\xi/2,\pm\sqrt{2\delta_1^2+3}/2)$. It follows from
this that the first two solutions approach each other with
increasing the parameter $\delta_1$ and merge at its critical
value, $\delta^\prime_{1}=1/\sqrt{2}$. For
$\delta_1>\delta^\prime_{1}$, the two electron pockets
corresponding to these merged solutions disappear (see Fig.~2c).
It should be noted that the total valley Chern number, $2\xi$,
remains constant with the disappearance of the two pockets.
Indeed, the central pocket of the bare bilayer graphene is
characterized by the Chern number $-\xi$, whereas each of the
three leg pockets has the Chern number $\xi$ (see, e.g.,
Ref.~\onlinecite{IRbil3}). Under the influence of the dressing
field, the central pocket ``annihilates'' with one of the leg
pockets, which conserves the total Chern number. Since the
``annihilation'' of the two pockets changes the Fermi topology
abruptly --- from the quadruply-connected Fermi surface to the
doubly-connected one --- the Lifshitz transition takes place. It
should be noted that it is similar to the Lifshitz transition in
the strained bilayer graphene~\cite{Falko1,Falko2}. Namely, the
dressing field linearly polarized along the $x$ axis effects on
BLG similarly to an uniform  strain applied along the same axis.
As to experimental manifestation of the Lifshitz transition in
bilayer graphene, it leads to the pronounced modification of the
Landau levels in the presence of a magnetic field~\cite{Falko1}.
If the magnetic field, $B$, is weak enough, the inverse magnetic
length, $\sqrt{eB/\hbar}$, is less than the distance between the
central electron pocket and the additional pockets pictured in
Fig.~2 . In this case, electronic states in the different pockets
are quantized by the magnetic field independently and the ground
electronic state is 16-fold degenerated. As a consequence, the
filling factor $\nu=16$ appears in the quantum Hall conductance.
After the Lifshitz transition, only two of four electron pockets
survive. This lifts the additional degeneracy and the filling
factor $\nu=8$ will be observed in the Hall measurements. Thus,
the Lifshitz transition leads to the possibility of optical
control of quantum Hall effect, what can be observed in
magnetotransport experiments.
\begin{figure}
\includegraphics[width=1.0\columnwidth]{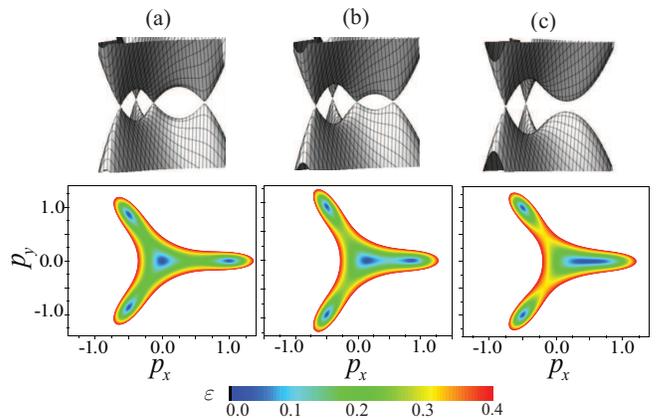}
\caption{\label{fig:figure2} (Color online) The three-dimensional
plots of the electron energy spectrum of irradiated AB-stacked
bilayer graphene, $\varepsilon$, near the band edge (top) and the
corresponding equi-energy maps (bottom) for the linearly polarized
dressing field with the photon energy $\hbar\omega=25$~meV and
different irradiation intensities, $I$: (a) $I=0$~kW/cm$^2$, (b)
$I=10$~kW/cm$^2$, (c) $I=45$~kW/cm$^2$.}
\end{figure}

In the case of circularly polarized dressing field
($\theta=\pi/4$), the effective Hamiltonian (\ref{Heff}) reads
\begin{align}\label{HC}
\hat{{\cal H}}_{\mathrm{eff}}=\hat{{\cal
H}}_{AB}+\frac{\xi\delta_1^2\delta_2}{2}(1-4p^2)\sigma_z+O(\delta_1^2\delta_2^2)
\end{align}
and results in the dispersion equation,
\begin{align}\label{e4}
\varepsilon^2=\frac{\varepsilon_g^2}{4}+p^2(1-\varepsilon_g^2)-2p^3\cos
3\phi +p^4(1+2\varepsilon_g^2),
\end{align}
where $\varepsilon_g=\sqrt{2}\delta_1^2\delta_2$ is the
field-induced energy gap at the $K_\pm$ point. Opening the energy
gap, the circularly polarized field effects on BLG similarly to
the interlayer asymmetry induced by the gate
voltage~\cite{Falko2}. For the critical value of the energy gap,
$\varepsilon_g=1$, the term $\sim p^2$ in Eq.~(\ref{e4}) vanishes
and the electron dispersion is $\varepsilon=\pm\sqrt{0.25-2p^3\cos
3\phi}$, where $\phi$ is the polar angle in the BLG plane. At the
Fermi energy $\varepsilon_F=1/2$, this dispersion corresponds to
the Fermi surface shown in Fig.~3a. The structure of the Fermi
surface for different irradiation intensities, $I$, is shown in
the phase diagram 3b. The critical point of the phase diagram ---
so-called monkey-saddle point~\cite{Shtyk2017} --- corresponds to
the critical energy gap, $\varepsilon_g=1$. At this point, the
Fermi surface is unstable with respect to small variations of both
the Fermi energy and the irradiation intensity: It undergoes one
of the four possible Lifshitz transitions depicted in Fig.~3b.
Similarly to the case of linear polarization, the Lifshitz
transition induced by the circularly polarized dressing field will
manifest itself in magnetoconductance. Namely, the three separated
pockets at the Fermi surface will lead to the filling factor
$\nu=12$ in the corresponding steps of quantum Hall conductivity.
Moreover, it should be noted that the critical point pictured in
Fig.~3b is characterized by the diverging density of states (DoS),
which is similar to the conventional Van Hove singularity. Indeed,
the density of states, $\rho$, in the vicinity of the critical
point reads
\begin{align}\label{dos}
&\rho(\varepsilon)=\frac{1}{(2\pi)^2}\int d^2 \mathbf{p}\delta_1(\varepsilon-\sqrt{1/4-2p^2\cos 3\phi})=\nonumber\\
&=\frac{\sqrt{\varepsilon}}{9(2\pi)^2}\beta\left(\frac{1}{6},\frac{1}{2}\right)\left|\varepsilon^2-\frac{1}{4}\right|^{-1/3},
\end{align}
where $\beta(x,y)$ is the beta function (see the plot of the DoS
in Fig.~3c). Since this DoS singularity leads to the instabilities
of the system with respect to arbitrary weak
interactions~\cite{Shtyk2017}, which can be observed
experimentally.

It follows from the aforesaid that a dressing field can induce the
Lifshitz transitions in BLG, which depend strongly on the field
polarization. Though the Hamiltonians (\ref{HL}) and (\ref{HC})
were derived within the perturbation theory, the claimed
phenomenon is qualitatively the same for any strength of the
electron-field coupling. It should be stressed that the exact
numerical solution of the Floquet-Magnus problem (\ref{03}) leads
to slightly shifted critical points of the Lifshitz transitions
but does not affect their structure. As to experimental
observation of the optically induced Lifshitz transition, a source
of intense far-infrared radiation is required. Particularly, a
source of the dressing field with the photon energy 25 meV must
provide the output power of several mW in order to observe the
effects pictured in Fig.~3. This output power has been reported
for the quantum cascade lasers (see, e.g., Ref.~\onlinecite{QCL})
which look most appropriate for experimental observation of the
discussed effects. It should be noted also that the absorption of
the dressing field might provoke thermal fluctuations of electron
gas. As a consequence, differentiation between different
topological states can be complicated. In order to avoid this, an
experimental set-up must include a high-efficient thermostat.
Since the energy difference between different topological states
pictured in Fig.~3 is of meV scale, the thermal fluctuations
should be less than 1 K. Such a thermal stability can be provided
by state-of-the-art experimental equipment. It should be noted
also that the Fermi energy pictured in Fig.~3 can be modified
experimentally by introducing doping on the sample. However, we
have to take into account that an electromagnetic field can be
considered as a dressing field within the conventional Floquet
formalism if the collisional (Drude) absorption of the field by
conduction electrons can be neglected (see, e.g.,
Refs.~\onlinecite{Kibis_14,Morina_15}). To neglect the Drude
absorption of a high-frequency field in a doped sample, the
well-known condition, $\omega\tau\gg1$, should be satisfied. Since
the doping of the sample decreases the mean free time of
conduction electron, $\tau$, the density of doping impurities
should be small enough to satisfy this condition. Alternatively,
the gate voltage can be used to control the Fermi energy in
bilayer graphene (see, e.g., Ref.~\onlinecite{Falko2}).

\begin{figure}
\includegraphics[width=1.0\columnwidth]{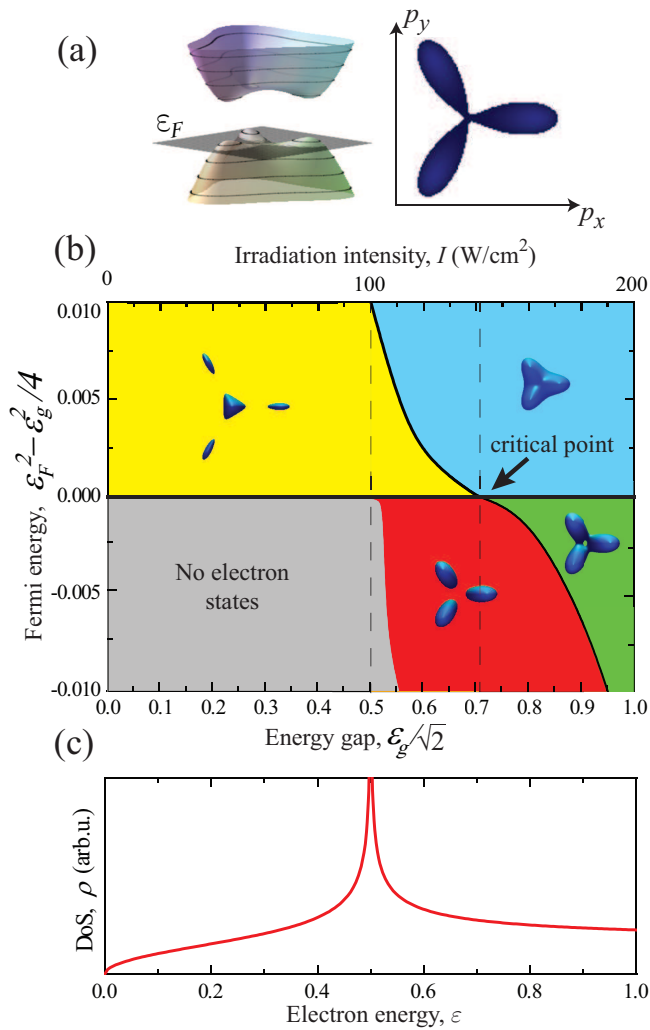}
\caption{\label{fig:figure3} (Color online) The electronic
structure of irradiated AB-stacked bilayer graphene for the
circularly polarized dressing field with the photon energy
$\hbar\omega=25$~meV: (a) Fermi surface at the monkey-saddle
point; (b) Phase diagram of the Fermi surface topology; (c)
Density of states spectrum near the monkey-saddle point.}
\end{figure}

\section{Conclusion and acknowledgements}

We demonstrated theoretically that a strong high-frequency
electromagnetic field (dressing field) can be used as an effective
tool to control topology of the Fermi surface of bilayer graphene
(BLG). The discussed Lifshitz transition strongly depends on the
polarization of the dressing field. Namely, the linearly polarized
field can induce the Lifshitz transition from the
quadruply-connected Fermi surface to the doubly-connected one,
whereas the circularly polarized field induces the multicritical
point, where the four different Fermi topologies may coexist.
Physically, the linearly polarized field effects on BLG similarly
to an uniform mechanical strain applied in the BLG plain along the
polarization vector, whereas the circularly polarized field
effects similarly to a gate voltage inducing the energy gap.
Therefore, the optically induced Lifshitz transition in BLG can be
used as a tool to control electronic properties of BLG-based
structures.

The work was partially supported by RISE Program (project CoExAN),
FP7 ITN Program (project NOTEDEV), Russian Foundation for Basic
Research (project 17-02-00053), Rannis project 163082-051, and
Ministry of Education and Science of Russian Federation (projects
3.4573.2017/6.7, 3.2614.2017/4.6, 3.1365.2017/4.6, 3.8884.2017/8.9
and 14.Y26.31.0015).

\end{document}